\documentclass[preprint]{jpsj2}
\title{Collective Excitations of Bose-Einstein Condensates \\
in a Double-Well Potential}

\author{Ippei \textsc{Danshita}\thanks{E-mail address: danshita@kh.phys.waseda.ac.jp},
Kyota \textsc{Egawa},
Nobuhiko \textsc{Yokoshi}
and Susumu \textsc{Kurihara}
}

\inst{
Department of Physics, Waseda University, \=Okubo, Shinjuku, Tokyo 169-8555.
}

\abst{We investigate collective excitations of Bose-Einstein condensates at absolute zero in a double-well trap.
We solve the Bogoliubov equations with a double-well trap, and show that the crossover from the dipole mode to the Josephson plasma mode occurs in the lowest energy excitation.
It is found that the anomalous tunneling property of low energy excitations is crucial to the crossover.}

\kword{Bose-Einstein condensate, Bogoliubov excitation, double-well trap, Josephson effect, anomalous tunneling}

\begin{document}
\maketitle

\section{Introduction} 
One of the most intriguing features in Bose-condensed systems is the coherence of the macroscopic quantum phase of the order parameter.
A Bose-Einstein condensate behaves as a quantum mechanical wave with a macroscopic quantum phase.
The macroscopic quantum phase difference of a condensate trapped in a double-well potential was measured by means of the matter wave interference both in destructive~\cite{rf:interfere} and {\it non}destructive~\cite{rf:nondestructive} ways, and the coherence of the Bose-Einstein condensate was clearly verified.

The Josephson effect, which was first predicted for superconductor tunnel junctions~\cite{rf:joseph_origin}, also clearly exhibits the coherence of a Bose-Einstein condensate.
The two mode approximation is useful to describe the Josephson-like dynamics of a weakly linked Bose-Einstein condensate in a double-well potential (see, e.g., refs. 4-7 and references therein).
Using the two mode approximation, various types of the Josephson-like dynamics have been predicted, which include Josephson plasma oscillation and nonlinear self-trapping.~\cite{rf:smerzi,rf:zwerg,rf:BEC}
Recently, the Josephson plasma oscillation and the nonlinear self-trapping have been observed by Albiez {\it et al.} in a condensate of atomic gases trapped in a double-well potential.~\cite{rf:JPST}
Because of the high tunability of system parameters, such a system provides an ideal stage for observing the Josephson-like dynamics.

Since the Josephson plasma mode is a small amplitude oscillation from a static equilibrium, it can be treated as a kind of Bogoliubov excitations.~\cite{rf:parao}
When the potential barrier separating the condensate does not exist, the lowest energy excitation is the dipole mode.
In contrast, when the potential barrier is so strong that the two mode approximation can be valid, the lowest excitation energy accords with the Josephson plasma energy.
In other words, the crossover from the dipole mode to the Josephson plasma mode occurs in the lowest energy excitation.

Salasnich {\it et al}. numerically solved the Bogoliubov equations in a double-well trap with a harmonic confinement and a Gaussian potential barrier to obtain the collective excitation energies.~\cite{rf:numel_dw}
They found that the lowest excitation energy decreases as the barrier strength increases; however they did not refer to the relation between the reduction of the lowest excitation energy and the crossover to the Josephson plasma mode.
The excitation energy in a double-well potential can be determined by a solution of a scattering problem of Bogoliubov excitations through the potential barrier used for preparing the double-well potential.
Kagan {\it et al.} studied the scattering problem of Bogoliubov excitations, and predicted that a potential barrier is transparent for low energy excitations, which is called {\it anomalous tunneling}.~\cite{rf:antun}
It is expected that we can relate the anomalous tunneling to the crossover to the Josephson plasma mode, when the Josephson plasma mode is regarded as a Bogoliubov excitation.

In the present paper, we study collective excitations of Bose-Einstein condensates at temperature $T=0$ in a double-well potential, and discuss the crossover from the dipole mode to the Josephson plasma mode.
Adopting a box-shaped double-well potential, we analytically calculate the lowest excitation energy and the Josephson plasma energy for arbitrary values of the barrier strength.
We show that the lowest excitation energy asymptotically approaches the Josephson plasma energy as the barrier strength increases.
We also find that the anomalous tunneling determines the region of the barrier strength where the crossover occurs.
Moreover, we numerically calculate the lowest excitation energy in case of a double-well potential consisting of a harmonic confinement and a Gaussian potential barrier.
It is shown that the mechanism of the crossover is valid also in this experimentally accessible trap.

The present paper is organized as follows.
In \S \ref{sec:cond}, solving the Gross-Pitaevskii equation with a box-shaped double-well potential, we analytically calculate the condensate wave function and the Josephson plasma energy.
In \S \ref{sec:exci}, solving the Bogoliubov equations with the box-shaped double-well potential, we analytically obtain the excitation spectra of condensates.
In \S \ref{sec:cross}, focusing our attention on the lowest energy excitation, we discuss the crossover from the dipole mode to the Josephson plasma mode.
In \S \ref{sec:numel}, we numerically calculate the lowest excitation energy for a condensate in an experimentally accessible double-well potential.
In \S \ref{sec:conc}, we summarize our results.
\section{Condensate Wave Functions and Josephson Plasma Energy in a Box-Shaped Double-Well Potential}\label{sec:cond}
We first consider a Bose-Einstein condensate in a box-shaped trap which consists of a radial harmonic confinement and end caps in the axial direction (the $x$ axis).
We assume that the frequency $\omega_{\perp}$ of the radial harmonic potential is large enough compared to the excitation energy for the axial direction.
Then, one can justify the one-dimensional treatment of the problem.
Such a configuration was realized in a recent experiment.~\cite{rf:box}
Setting a potential barrier in the center of the trap, a double-well potential is created.
Adopting rigid walls and a $\delta$-function potential barrier as the end caps and the barrier respectively, the double-well potential is written as
  \begin{eqnarray}
    V(x)=\left\{
         \begin{array}{cc}
           V_0\delta(x),\,\,|x|<a,\\
           \infty,\,\,|x|\geq a,
         \end{array}\right.\label{eq:pote}
  \end{eqnarray}
where $a$ is the size of a well.

Since we are mainly interested in collective excitations of a double-well trapped condensate at the absolute zero temperature, our formulation of the problem is based on the Bogoliubov theory.
The Bogoliubov theory consists of the time-independent Gross-Pitaevskii equation and the Bogoliubov equations.~\cite{rf:BEC}
In this section, we solve the time-independent Gross-Pitaevskii equation:
  \begin{eqnarray}
    \Bigl[-\frac{\hbar^2}{2m}\frac{d^2}{dx^2}+V(x)+g|\Psi_0(x)|^2\Bigr]
    \Psi_0(x) = \mu\Psi_0(x),
    \label{eq:sGPE}
  \end{eqnarray}
and obtain the condensate wave function $\Psi_0(x)$.
Here $\mu$ is the chemical potential and $m$ is the mass of an atom.
Since the radial confinement is harmonic and sufficiently tight, the coupling constant is affected by the harmonic oscillator length $a_{\perp}$ of the radial confinement as $g=\frac{2\hbar^2a_s}{ma_{\perp}^2}$, where $a_s$ is the $s$-wave scattering length~\cite{rf:1dime}.
\begin{figure}[tb]
\begin{center}
\includegraphics[scale=0.2]{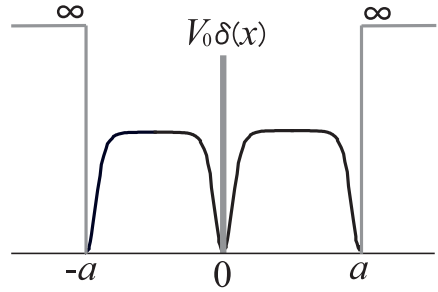}
\end{center}
\caption{\label{fig:box}
Schematic picture of a condensate in a box-shaped double-well potential.
Black solid line represents the condensate wave function.
                              }
\end{figure}

In \S \ref{sec:cross}, we will discuss the crossover from the dipole mode to the Josephson plasma mode by comparing the lowest excitation energy to the Josephson plasma energy.
Using solutions of the Gross-Pitaevskii equation, one can calculate the Josephson plasma energy.
In order to calculate the Josephson plasma energy, we need to obtain not only the lowest energy symmetric solution ($\Psi_0^{\mathrm{sm}}$) but also antisymmetric ($\Psi_0^{\mathrm{an}}$) solution of eq. \!(\ref{eq:sGPE}).
%
Assuming that the system size is much larger than the healing length $\xi=\frac{\hbar}{\sqrt{m\mu}}$, the condensate wave function near the center of each well is not affected by the potential barrier and the rigid wall.
Then, one can approximately obtain the symmetric and antisymmetric solutions of eq. \!(\ref{eq:sGPE}) as
 \begin{eqnarray}
  \Psi_0^{\mathrm{sm}}(x) &=& \sqrt{\frac{\mu_{\mathrm{sm}}}{g}}\times\Biggl\{
   \begin{array}{ll}
   \mathrm{tanh}\left(\frac{|x|+x_0}{\xi_{\mathrm{sm}}}\right),
   \,|x|<\frac{a}{2},
   \\
   \mathrm{tanh}\left(\frac{-|x|+a}{\xi_{\mathrm{sm}}}\right),
   \, \frac{a}{2}\leq |x|<a,
   \end{array}\Biggr.\label{eq:gs}
 \end{eqnarray}
 \begin{eqnarray}
  \Psi_0^{\mathrm{an}}(x) &=& \sqrt{\frac{\mu_{\mathrm{an}}}{g}}\times\left\{
   \begin{array}{ll}
   \mathrm{tanh}\left(\frac{x}{\xi_{\mathrm{an}}}\right),
   \,\,\,\,|x|<\frac{a}{2},
   \\
   \mathrm{sgn}(x)\,\mathrm{tanh}\left(\frac{-|x|+a}{\xi_{\mathrm{an}}}\right),
   \frac{a}{2}\leq |x|<a,
   \end{array}\right.\label{eq:ex}
  \end{eqnarray}
where $\mu_{\mathrm{sm}(\mathrm{an})}$ and $\xi_{{\rm sm}({\rm an})}$ is the chemical potential and the healing length of the symmetric (antisymmetric) state.
A constant $x_0$ reflects value of the condensate wave function at $x=0$, which is determined by the boundary conditions
  \begin{eqnarray}
  \Psi_0(-0)=\Psi_0(+0), \label{eq:bound0}
  \end{eqnarray}
  \begin{eqnarray}
  \frac{d\Psi_0}{dx}\biggl.\biggr|_{x=-0}
  &=&
  \frac{d\Psi_0}{dx}\biggl.\biggr|_{x=+0}-\frac{2mV_0}{\hbar^2}\Psi_0(0),
  \label{eq:bound0_de}
  \end{eqnarray}
as
  \begin{eqnarray}
  \mathrm{tanh}\frac{x_0}{\xi_{\mathrm{sm}}}=
  \frac{-V_0+\sqrt{V_0^2+4(\mu_{\mathrm{sm}}\xi_{\mathrm{sm}})^2}}
  {2\mu_{\mathrm{sm}}\xi_{\mathrm{sm}}}.
  \end{eqnarray}

We next calculate the Josephson plasma energy $\varepsilon_{\mathrm{JP}}=\sqrt{E_{\mathrm{J}}E_{\mathrm{C}}}$, which is easily derived from the Josephson Hamiltonian:
  \begin{eqnarray}
  H_{\mathrm{J}}=\frac{E_{\mathrm{C}}k^2}{2}-E_{\mathrm{J}}\,\mathrm{cos}\varphi,
  \end{eqnarray}
where $k$ and $\varphi$ represent the population difference and the phase difference between the two wells.~\cite{rf:BEC}
The Josephson coupling energy $E_{\mathrm{J}}$ expresses the overlapping integral of condensate wave functions in the two wells, and the capacitive energy $E_{\mathrm{C}}$ is proportional to inverse of the compressibility.
They are defined as
  \begin{eqnarray}
      E_{\mathrm{J}}&=&\frac{E_{\mathrm{an}}-E_{\mathrm{sm}}}{2},
      \label{eq:jose_J}\\
      E_{\mathrm{C}}&=&4\frac{d\mu_{\mathrm{sm}}}{dN_0},\label{eq:jose_C}
  \end{eqnarray}
where $E_{\mathrm{sm(an)}}$ is the mean field energy of the symmetric (antisymmetric) state.~\cite{rf:BEC}
Hence, we need to calculate the chemical potential and the mean field energy in order to obtain the Josephson plasma energy.

\begin{figure}[tbp]
\begin{center}
\includegraphics[scale=0.25]{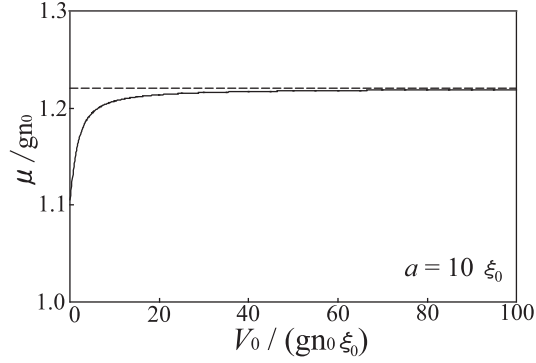}
\end{center}
\caption{\label{fig:chem}
Chemical potentials $\mu_{\mathrm{sm}}$ (solid line) and $\mu_{\mathrm{an}}$ (dashed line) as functions of the potential strength $V_0$ are shown.}
\end{figure}

The chemical potential is related to the number $N_0$ of condensate atoms by the normalization condition:
  \begin{equation}
     \int_{-a}^{a}dx|\Psi_0(x)|^2=N_0.\label{eq:norm}
  \end{equation}
Substituting eqs. \!(\ref{eq:gs}) and (\ref{eq:ex}) into the normalization condition, we derive the relations between the chemical potentials and $N_0$:
  \begin{eqnarray}
    \frac{\mu_{\mathrm{sm}}}{gn_0}\left(1-2\frac{\xi_{\mathrm{sm}}}{a}+
    \frac{\xi_{\mathrm{sm}}}{a}\mathrm{tanh}\frac{x_0}{\xi_{\mathrm{sm}}}
    \right)=1,\label{eq:gschem}
  \end{eqnarray}
  \begin{eqnarray}
    \frac{\mu_{\mathrm{an}}}{gn_0}\left(1-2\frac{\xi_{\mathrm{an}}}{a}
    \right)=1,
    \label{eq:exchem}
  \end{eqnarray}
where $n_0\equiv \frac{N_0}{2a}$ is the averaged density of the condensate.
One can obtain approximate solutions of eq. \!(\ref{eq:gschem}) in the limits of $V_0 \gg gn_0\xi_0$ and $V_0 \ll gn_0\xi_0$, where $\xi_0\equiv\frac{\hbar}{\sqrt{mgn_0}}$.
When $V_0 \gg gn_0\xi_0$, we expand eq. \!(\ref{eq:gschem}) into power series of $\frac{\xi_0}{a}$ and $\frac{gn_0\xi_0}{V_0}$, and obtain
  \begin{eqnarray}
    \frac{\mu_{\mathrm{sm}}}{gn_0} \simeq
     1+\frac{2\xi_0}{a}+\frac{2\xi_0^2}{a^2}-\frac{gn_0\xi_0^2}{aV_0}
     +\frac{\xi_0^3}{a^3}-\frac{3gn_0\xi_0^3}{a^2 V_0}. 
     \label{eq:cheml_dw}
  \end{eqnarray}
In a similar way, when $V_0 \ll gn_0\xi_0$, we expand eq. \!(\ref{eq:gschem}) into power series of $\frac{\xi_0}{a}$ and $\frac{V_0}{gn_0\xi_0}$, and obtain
  \begin{eqnarray}
    \frac{\mu_{\mathrm{sm}}}{gn_0} \simeq
    1+\frac{\xi_0}{a}+\frac{\xi_0^2}{2a^2}+\frac{V_0}{2agn_0}
    +\frac{\xi_0^3}{8a^3}+\frac{\xi_0 V_0}{4a^2 gn_0}
    -\frac{V_0^2}{8a\xi_0(gn_0)^2}.
    \label{eq:chems_dw}
  \end{eqnarray}
Meanwhile, the solution of eq. \!(\ref{eq:exchem}) is
  \begin{eqnarray}
    \frac{\mu_{\mathrm{an}}}{gn_0} 
    &=& 1+\frac{2\xi_0^2}{a^2}+2\frac{\xi_0}{a}\sqrt{1+\frac{\xi_0^2}{a^2}},
    \nonumber\\
    &\simeq&
    1+\frac{2\xi_0}{a}+\frac{2\xi_0^2}{a^2}+\frac{\xi_0^3}{a^3}.
    \label{eq:chemex_dw}
  \end{eqnarray}
In eqs. \!(\ref{eq:cheml_dw}), (\ref{eq:chems_dw}) and (\ref{eq:chemex_dw}), we express the expansions up to the third order of the small parameters.
In Fig. \ref{fig:chem}, we show the chemical potentials $\mu_{\mathrm{sm}}$ and $\mu_{\mathrm{an}}$ as functions of the potential strength $V_0$.
While $\mu_{\mathrm{an}}$ is constant, $\mu_{\mathrm{sm}}$ increases monotonically and approaches the value of $\mu_{\mathrm{an}}$ as $V_0$ increases.
This means that the symmetric state and the antisymmetric state are degenerate in the strong potential limit $V_0\to\infty$.

The mean field energy $E$ can be calculated using the expression
  \begin{eqnarray}
   E=
   \int_{-a}^{a}dx\Psi_0^{\ast}(x)
   \Bigl[-\frac{\hbar^2}{2m}\frac{d^2}{dx^2}
   +V(x)+\frac{g}{2}|\Psi_0(x)|^2\Bigr]\Psi_0(x).
  \end{eqnarray}
In the limit of $V_0\gg gn_0\xi_0$, one can approximately obtain
  \begin{eqnarray}
   \frac{E_{\mathrm{sm}}}{N_0 gn_0}&\simeq&
     1+\frac{4\xi_0}{3a}+\frac{2\xi_0^2}{a^2}
     -\frac{gn_0\xi_0^2}{2aV_0}+\frac{2\xi_0^3}{a^3}
     -\frac{2gn_0\xi_0^3}{a^2 V_0},\label{eq:mean_gs}\\
   \frac{E_{\mathrm{an}}}{N_0 gn_0}&\simeq&
     1+\frac{4\xi_0}{3a}+\frac{2\xi_0^2}{a^2}+\frac{2\xi_0^3}{a^3},
     \label{eq:mean_ex}
  \end{eqnarray}
where $E_{\mathrm{sm}(\mathrm{an})}$ is the mean field energy of the symmetric (antisymmetric) state.
\begin{figure}[tbp]
\begin{center}
\includegraphics[scale=0.25]{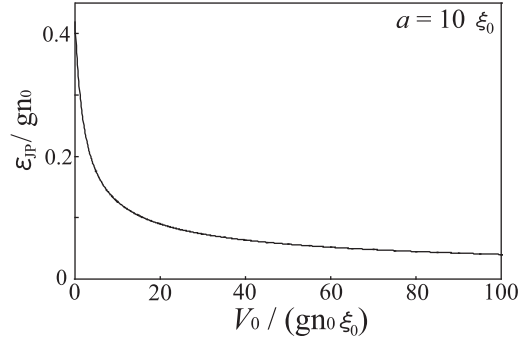}
\end{center}
\caption{\label{fig:JP_energy}
Josephson plasma energy $\varepsilon_{\mathrm{JP}}$ as a function of the potential strength $V_0$ is shown.}
\end{figure}

Substituting the chemical potential of eq. \!(\ref{eq:cheml_dw}) and the mean field energies of eqs. \!(\ref{eq:mean_gs}) and (\ref{eq:mean_ex}) into eqs. \!(\ref{eq:jose_J}) and (\ref{eq:jose_C}), one obtains
  \begin{eqnarray}
    \frac{E_{\mathrm{J}}}{gn_0}&\simeq&
    \frac{N_0gn_0\xi_0^2}{4V_0a}\left(1+\frac{4\xi_0}{a}\right),\\
    \frac{E_{\mathrm{C}}}{gn_0}&\simeq&
    \frac{4}{N_0}\left(1+\frac{\xi_0}{a}\right)
  \end{eqnarray}
when $V_0\gg gn_0\xi_0$.
These equations yield the Josephson plasma energy
  \begin{eqnarray}
    \frac{\varepsilon_{\mathrm{JP}}}{gn_0}\simeq
      \sqrt{\frac{gn_0\xi_0^2}{V_0a}}\left(1+\frac{5\xi_0}{2a}\right),
      V_0\gg gn_0\xi_0.\label{eq:josephp}
  \end{eqnarray}
The Josephson plasma energy is shown in Fig. \ref{fig:JP_energy}, as a function of the potential strength.
In \S \ref{sec:cross}, we will see that the lowest excitation energy coincides with this Josephson plasma energy for the sufficiently strong potential.

\section{Excitation Spectrum in a Box-Shaped Double-Well Potential}\label{sec:exci}
The aim of this section is to calculate the excitation spectra of condensates in a box-shaped double-well potential.
The excitations of a condensate correspond to the small fluctuations of the condensate wave function $\Psi_0(x,t)$ around the stationary solution of eq. \!(\ref{eq:sGPE}),
  \begin{eqnarray}
  \delta\Psi_0(x,t)=e^{-\frac{{\rm i}\mu t}{\hbar}}
  \sum_j\left(u_j(x)e^{-\frac{{\rm i}\varepsilon_j t}{\hbar}}
  -v_j^*(x) e^{\frac{{\rm i}\varepsilon_j t}{\hbar}}\right),
  \end{eqnarray}
where $\varepsilon_j$ is the energy of the excitation in an eigenstate labeled by $j$.
The wave functions $(u_j(x), v_j(x))^{\bf t}$ of the excitations fulfill the Bogoliubov equations~\cite{rf:BEC},
       \begin{eqnarray}
           \begin{array}{cc}
               \left(
                 \begin{array}{cc}
                 H_0 & -g{\Psi_0(x)}^2 \\
                 g\Psi_0^{\ast}(x)^2 & -H_0
                 \end{array}
               \right) 
               \left(
                 \begin{array}{cc}
                 u_j(x) \\ v_j(x)
                 \end{array}
               \right)
               = \varepsilon_j\left(
                 \begin{array}{cc}
                   u_j(x) \\ v_j(x)
                 \end{array}
               \right),               
           \end{array}\\ \label{eq:BdGE}
           H_0 = -\frac{\hbar^2}{2m}\frac{d^2}{dx^2}
               -\mu+V\left(x\right)+2g|\Psi_0(x)|^2.
       \end{eqnarray}
Solving these equations with the condensate wave function of eq. \!(\ref{eq:gs}), we shall obtain the excitation spectrum.
Hereafter the chemical potential $\mu$ denotes the chemical potential $\mu_{\rm sm}$ of the symmetric state.

In order to relate the problem to tunneling properties of the Bogoliubov excitation, we separate the system into three regions, or $-a < x \leq -\frac{a}{2}$, $|x| < \frac{a}{2}$ and $\frac{a}{2}\leq x<a$.
We solve the Bogoliubov equations analytically in each region, and derive the equation to determine the excitation spectrum by connecting each solution smoothly.

At first, we find solutions of the Bogoliubov equations in the central region ($|x| < \frac{a}{2}$).
There exist two independent solutions with a same excitation energy $\varepsilon$, corresponding to two types of scattering process.~\cite{rf:KP}
One solution $\psi^l(x)\equiv (u^l(x), v^l(x))^{\bf t}$ describes the process where a Bogoliubov excitation comes from left, and the other solution $\psi^r(x)\equiv (u^r(x), v^r(x))^{\bf t}$ describes the process where a Bogoliubov excitation comes from right.
The solution $\psi^l(x)$ is written as
  \begin{eqnarray}
     {\psi}^l(x)     
\!\!\!&=&\!\!\!\left\{\begin{array}{ll}
          \left(\!\begin{array}{cc}
                    u_1 \\ v_1
                         \end{array}\!\right) e^{{\rm i}p_+ x}
          +r\left(\!\begin{array}{cc}
                    u_2 \\ v_2
                         \end{array}\!\right) e^{-{\rm i}p_+ x}
          +b\left(\!\begin{array}{cc}
                    u_3 \\ v_3
                         \end{array}\!\right) e^{p_- x},
                         & \!\!\!x<0,  \\
          t\left(\!\begin{array}{cc}
                    u_1 \\ v_1
                         \end{array}\!\right) e^{{\rm i}p_+ x}
          +c\left(\!\begin{array}{cc}
                    u_4 \\ v_4
                         \end{array}\!\right) e^{-p_- x},
                         & \!\!\!x>0,
          \end{array}\right.\label{eq:lcs}
  \end{eqnarray}
where $p_{\pm}=\sqrt{\frac{2m}{\hbar^2}(\sqrt{\mu^2+\varepsilon^2}\mp\mu)}$ satisfies the Bogoliubov spectrum for a uniform system.~\cite{rf:antun,rf:KP,rf:kovri}
The coefficients $r$, $b$, $t$, and $c$ are the amplitudes of reflected, left localized, transmitted, and right localized components, respectively.
All the coefficients can be determined by the boundary conditions at $x=0$.
Thus, we obtained $\psi^l(x)$, and we can also find $\psi^r(x)$ in the same way.

Expanding $|t|$ and the phase $\delta$ of $t$ around $\varepsilon=0$,~\cite{rf:KP} one can analytically obtain
  \begin{eqnarray}
    |t| &\simeq& 1-\alpha\left(\frac{\varepsilon}{\mu}\right)^2,
    \label{eq:unexp_pr}\\ 
    \delta &\simeq& \beta\frac{\varepsilon}{\mu}.
    \label{eq:tnexp_ph}
  \end{eqnarray}
The coefficients $\alpha$ and $\beta$ are
  \begin{eqnarray}
  \alpha=
  \frac{2(V_0 \!-\! \mu\xi)(V_0^3\!+\! \nu V_0^2\!+\! 2\nu(\mu\xi)^2
  - 4(\mu\xi)^3)+ 9(\mu\xi V_0)^2}{8(\mu\xi\nu)^2},
  \end{eqnarray}
  \begin{eqnarray}
  \beta=\frac{V_0^2+\nu V_0-3\mu\xi\nu+6(\mu\xi)^2}{2\mu\xi\nu},
  \end{eqnarray}
where
  \begin{eqnarray}
  \nu=\sqrt{V_0^2+4(\mu\xi)^2}.
  \end{eqnarray}
It is obvious from eqs. \!(\ref{eq:unexp_pr}) and (\ref{eq:tnexp_ph}) that the transmission coefficient $|t|^2$ approaches unity and the phase shift of $t$ approaches zero as the energy is reduced to zero; Kagan $\mathit{et \,\, al.}$ called this behavior {\it anomalous tunneling}~\cite{rf:antun}.
This expression is adequate for arbitrary values of the potential strength $V_0$.
Furthermore, when $\varepsilon \ll \mu$ and $V_0 \gg \mu\xi$, one can approximately obtain the reflection and transmission amplitudes,~\cite{rf:KP}
  \begin{eqnarray}
       r &=& \frac{\varepsilon V_0-\varepsilon\mu\xi
             +i\frac{\varepsilon^2 V_0}{\mu}}
             {\varepsilon V_0-\varepsilon\mu\xi+i\mu^2\xi},\label{eq:cfrf}
               \\
       t &=& \frac{\varepsilon\mu\xi+i\mu^2\xi}
       {\varepsilon V_0-\varepsilon\mu\xi+i\mu^2\xi}. \,
       \label{eq:cfct}
  \end{eqnarray}
In eq. (\ref{eq:cfct}), we can see that the peak around $\varepsilon=0$ in $|t|^2$ has the Lorentzian shape with half width $\Delta\varepsilon\sim \frac{\mu^2\xi}{V_0}$.
Such an anomalous tunneling property of the Bogoliubov excitations essentially affects the crossover from the dipole mode to the Josephson plasma mode of the lowest energy excitation, as discussed in the next section.

We can write a general solution of the Bogoliubov equations in the central region as a linear combination of $\psi^l(x)$ and $\psi^r(x)$:
  \begin{eqnarray}
    \psi(x)=\eta\psi^l(x)+\zeta\psi^r(x), \, |x|<\frac{a}{2}.
    \label{eq:centre}
  \end{eqnarray}
  
One can obtain an analytical solution of the Bogoliubov equations also in the left and right side regions ($\frac{a}{2}\leq |x| <a$).
It is
 \begin{eqnarray}
    u^{\mathrm{side}}(x)&=&
    \sqrt{\frac{2\mu}{\varepsilon}}
    \Biggl[\Biggr.
    \left(1+\frac{(p_+\xi)^2\mu}{2\varepsilon}
    \right)
    \mathrm{tanh}\left(\frac{|x|-a}{\xi}\right)
    \mathrm{cos}p_+(|x|-a)
    \nonumber\\
    &&+\frac{p_+\xi\mu}{2\varepsilon}
    \Biggl(\Biggr.\frac{(p_+\xi)^2}{2}+1
    -\mathrm{tanh}^2\left(\frac{|x|-a}{\xi}\right)
    +\frac{\varepsilon}{\mu}\Biggl.\Biggr)
    \mathrm{sin}p_+(|x|-a)
    \Biggl.\Biggr],\label{eq:sideu}
    \\
    v^{\mathrm{side}}(x)&=&
    \sqrt{\frac{2\mu}{\varepsilon}}
    \Biggl[\Biggr.
    \left(1-\frac{(p_+\xi)^2\mu}{2\varepsilon}
    \right)
    \mathrm{tanh}\left(\frac{|x|-a}{\xi}\right)
    \mathrm{cos}p_+(|x|-a)\nonumber\\
    &&-\frac{p_+\xi\mu}{2\varepsilon}
    \Biggl(\Biggr.\frac{(p_+\xi)^2}{2}+1
    -\mathrm{tanh}^2\left(\frac{|x|-a}{\xi}\right)
    -\frac{\varepsilon}{\mu}\Biggl.\Biggr)
    \mathrm{sin}p_+(|x|-a)
    \Biggl.\Biggr].\label{eq:sidev}
  \end{eqnarray}

Using the solutions of eqs. \!(\ref{eq:centre}), (\ref{eq:sideu}) and (\ref{eq:sidev}), one can construct solutions in all regions:
  \begin{eqnarray}
     {\psi}(x)
  \!\!\!&=&\!\!\!\left\{\begin{array}{ll}
                    F \psi^{\mathrm{side}}(x)
                         & \!\!\!-a<x\leq-\frac{a}{2},\\
                         \vspace{-2mm}
                         & \\
                         
                    \eta\psi^l(x)
                   +\zeta\psi^r(x)
                    & |x|<\frac{a}{2},\\\vspace{-2mm}
                    & \\
                    
                    G \psi^{\mathrm{side}}(x)
                         & \!\!\!\frac{a}{2}\leq x < a,
          \end{array}\right.\label{eq:spose}
  \end{eqnarray}
where $\eta=\zeta$ and $F=G$ are satisfied for collective excitations with even parity or $\eta=-\zeta$ and $F=-G$ for those with odd parity.
By imposing the boundary condition at $|x|=\frac{a}{2}$, the solutions can be smoothly connected.
As a result, one obtains the equation to determine the excitation spectrum:
  \begin{equation}
    (r \pm t)\mathrm{exp}\left({\rm i}(2p_+a+\gamma)\right)=1,
    \label{eq:eq_bspe}
  \end{equation}
where
  \begin{eqnarray}
  \gamma \equiv \mathrm{tan}^{-1}
  \left(\frac{-p_+\xi_{\mathrm{sm}}}{1-\frac{1}{4}(p_+\xi_{\mathrm{sm}})^2}
  \right).
  \end{eqnarray}
The positive (negative) sign in the left-hand side of eq. \!(\ref{eq:eq_bspe}) represents the excitations with even (odd) parity.
One can see from eq. \!(\ref{eq:eq_bspe}) that tunneling properties of the Bogoliubov excitations through the potential barrier affect the excitation spectrum explicitly.
Since the reflection and transmission amplitudes satisfy relations:~\cite{rf:KP}
  \begin{eqnarray}
    |t|^2+|r|^2=1,\label{eq:efcl}
  \end{eqnarray}
and
  \begin{eqnarray}
    t=|t|e^{{\rm i}\delta}, \,r=\pm {\rm i}|r|e^{{\rm i}\delta},\label{eq:prim}
  \end{eqnarray}
one obtains
  \begin{eqnarray}
   |r \pm t|=1.\label{eq:reftra}
  \end{eqnarray}
Using eq. \!(\ref{eq:reftra}), one can rewrite eq. \!(\ref{eq:eq_bspe}) in terms of the phase of the left-hand side as
  \begin{equation}
    2p_+ a+\gamma+\phi_{\pm}=2\pi n, \label{eq:ph_exp}
  \end{equation}
where $\phi_{\pm}$ is the phase of $r \pm t$.
\begin{figure}[tbp]
\begin{center}
\includegraphics[scale=0.25]{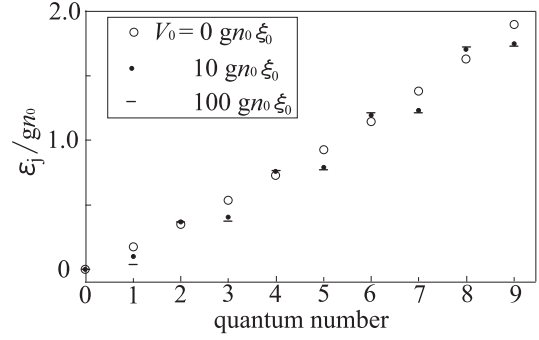}
\end{center}
\caption{\label{fig:epsart5}
Excitation spectra in the double-well traps with $V_0=0$, $V_0=10$ and $V_0=100$ are shown, where $a=10\xi_0$.
                  }
\end{figure}

Solving eq. \!(\ref{eq:ph_exp}) for excitation energy $\varepsilon$, we obtain the excitation spectra which consist of discrete energy levels as shown in Fig. \ref{fig:epsart5}.
Parity of quantum numbers expresses parity of the wave function of the collective excitations.
In the strong potential limit $V_0\rightarrow \infty$, the excitations labeled by $2l$ and $2l+1$ are degenerate, because the condensate is completely divided into each well.
Hence, as the potential barrier becomes stronger, energy of the even parity excitation labeled by $2l$ increases and energy of the odd parity excitation labeled by $2l+1$ decreases so that both energies approach.
The change of energies of the odd parity excitations is more pronounced than that of the even parity excitations, and such a tendency is qualitatively consistent with the numerical results in the case of the double-well traps with harmonic confinement~\cite{rf:numel_dw}.
Moreover, we see that the less the excitation energy is, the slower the change according to growth of the potential barrier is.
This implies that the anomalous tunneling behavior of low energy excitations buffers the effect of the potential barrier.

\section{Crossover from Dipole Mode to Josephson Plasma Mode}\label{sec:cross}
\begin{figure}[tbp]
\begin{center}
\includegraphics[width=3 in, height=2 in]{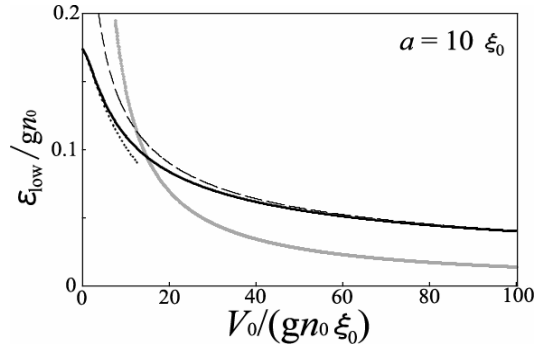}
\end{center}
\caption{\label{fig:lowest_ex}
Black solid line is the lowest excitation energy for $a=10\xi_0$ as a function of $V_0$.
Dotted line is the corrected dipole mode energy of eq. \!(\ref{eq:small_cor}), while the dashed line is the Josephson plasma energy shown in Fig. \ref{fig:JP_energy}.
Gray solid line is the half-width of the peak of the transmission coefficient.}
\end{figure}

In this section, we shall discuss the effect of the anomalous tunneling on the lowest energy excitation.
Solving eq. \!(\ref{eq:ph_exp}) with $n=0$ and odd parity, we obtain the lowest excitation energy as a function of the potential strength, which is represented by the solid line in Fig. \ref{fig:lowest_ex}.
In order to elucidate the effect of the anomalous tunneling on the lowest excitation energy $\varepsilon_{\mathrm{low}}$, we calculate it analytically in two limits.

On the one hand, when $\frac{gn_0\xi_0}{V_0} \gg \frac{\xi_0}{a}$, substituting the low energy expansion of the transmission amplitude of eqs. \!(\ref{eq:unexp_pr}) and (\ref{eq:tnexp_ph}), one obtains
  \begin{eqnarray}
    \frac{\varepsilon_{\mathrm{low}}}{gn_0}
    \simeq\frac{\pi\xi_0}{2a}\Biggl[\Biggr.
    1+\frac{\xi_0}{a}\biggl(\biggr.\frac{3}{2}
    -\frac{\beta}{2}-\sqrt{\frac{\alpha}{2}}
    -\frac{\sqrt{V_0^2+4(gn_0\xi_0)^2}-V_0}{4gn_0\xi_0}
    \biggl.\biggr)\Biggl.\Biggr],\label{eq:small_cor}
  \end{eqnarray}
which is plotted in Fig. \ref{fig:lowest_ex} by the dotted line.
Obviously, the influence of the potential barrier on the lowest excitation is relatively small, because the correction of $\varepsilon_{\rm low}$ due to the potential barrier is included in the small term of $O(\frac{\xi_0}{a})$ in eq. \!(\ref{eq:small_cor}).
On the other hand, when $\frac{gn_0\xi_0}{V_0} \ll \frac{\xi_0}{a}$,
one can obtain an approximate expression of the lowest excitation energy from eqs. \!(\ref{eq:cfrf}) and (\ref{eq:cfct}),
  \begin{eqnarray}
    \frac{\varepsilon_{\mathrm{low}}}{gn_0}\simeq
    \sqrt{\frac{gn_0 \xi_0^2}{V_0 a}}
          \left(1+\frac{5\xi_0}{2a}\right).
  \end{eqnarray}
This energy coincides with the Josephson plasma energy of eq. \!(\ref{eq:josephp}), and this result explicitly justifies treatment of the Josephson plasma oscillation by the two mode approximation for a sufficiently strong potential barrier.
In Fig. \ref{fig:lowest_ex}, one can see that the crossover from the dipole mode to the Josephson plasma mode occurs.
%
%
%
%
%
%
%

It is crucial to understand the physics of the crossover that two energy scales of, $\frac{gn_0\xi_0}{a}$ and $\frac{(gn_0)^2\xi_0}{V_0}$, determine whether the lowest energy excitation is the Josephson plasma mode or the dipole mode.
The former is comparable to the dipole mode energy.
The latter is comparable to the half-width $\Delta\varepsilon$ of the peak of the transmission coefficient $|t|^2$.
In other words, the potential barrier is almost transparent for the excitations with energy less than $\frac{(gn_0)^2\xi_0}{V_0}$.
The crossover is determined by whether the anomalous tunneling is effective or not for the lowest energy excitation.
When $\frac{\xi_0}{a} \ll \frac{gn_0\xi_0}{V_0}$, the potential barrier is almost transparent for the lowest energy excitation due to the anomalous tunneling; , the dipole mode is hardly affected by the potential barrier.
As the potential strength $V_0$ increases, contribution of the anomalous tunneling diminishes gradually and the lowest energy excitation comes to be significantly affected by the potential barrier
Consequently, when $\frac{\xi_0}{a} \gg \frac{gn_0\xi_0}{V_0}$, the lowest energy excitation becomes the Josephson plasma mode.
Figure \ref{fig:lowest_ex}, where the half-width of the anomalous tunneling (gray line) divides the lowest excitation energy into regions of the dipole mode and the Josephson plasma mode, clearly confirms such an interpretation of the crossover.

\section{Realistic Double-Well Potential}\label{sec:numel}
In the preceding sections, we considered the case in the double-well potential of eq. \!(\ref{eq:pote}) with rigid walls and a $\delta$-function potential barrier.
In this section, we numerically solve the Gross-Pitaevskii equation and the Bogoliubov equations in case of a realistic double-well potential.

Considering a double-well potential consisting of the magnetic trap and the blue-detuned laser beam, which is realized in the experiment of ref. 1, we adopt a double-well potential with a harmonic confinement and a Gaussian potential barrier:
  \begin{eqnarray}
  V(x)=\frac{m\omega_x^2 x^2}{2}+U_0\,{\rm exp}\left(-\frac{x^2}{\sigma^2}\right),
  \label{eq:realpote}
  \end{eqnarray}
where $\omega_x$ is the frequency of the harmonic potential.
The height $U_0$ and the width $\sigma$ of the potential barrier can be controlled in experiments by varying the intensity and the aperture of the laser beam, respectively.

With use of the Bogoliubov theory, one can discuss the crossover to the Josephson plasma mode in the same way as in the case of the box-shaped double-well potential.
We can numerically find the lowest symmetric and antisymmetric solutions of the Gross-Pitaevskii equation with the double-well potential of eq. \!(\ref{eq:realpote}); accordingly we can calculate the Josephson coupling energy $E_{\rm J}$ and the capacitive energy $E_{\rm C}$ using the expressions of eq. \!(\ref{eq:jose_J}) and eq. \!(\ref{eq:jose_C}).
As a result, we can obtain the Josephson plasma energy $\varepsilon_{\rm JP}$.
Furthermore, substituting the lowest symmetric solution of the Gross-Pitaevskii 
equation into the Bogoliubov equations, we can numerically calculate the lowest excitation energy.

\begin{figure}[tb]
\begin{center}
\includegraphics[scale=0.25]{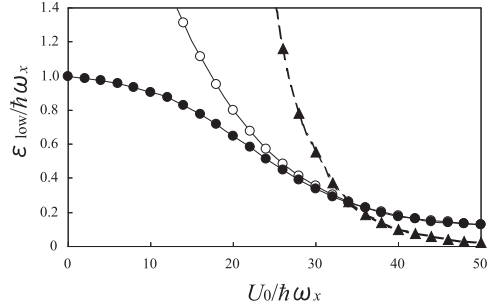}
\end{center}
\caption{Closed circles represent the lowest excitation energy, and open circles represent the Josephson plasma energy. Triangles represent the half-width of the peak of the transmission coefficient. We consider a condensate of $^{23}{\rm Na}$ atoms. The values of parameters are as follows; the number of the condensed atoms is $N_0=3000$, the frequency of the radial confinement is $\omega_{\perp}=250\times 2\pi \,\,{\rm Hz}$, the frequency of the axial confinement is $\omega_{\perp}=10\times 2\pi \,\,{\rm Hz}$, the {\it s}-wave scattering length is $a_s=3 \,{\rm nm}$ and the width of the potential barrier is $\sigma=1.33 \,{\rm \mu m}$.}
\label{fig:harmo}
\end{figure}

Open circles and closed circles in Fig. \ref{fig:harmo} represent the lowest excitation energy and the Josephson plasma energy, respectively.
In Fig. \ref{fig:harmo} one can see that the lowest excitation energy reduces as height of the potential barrier increases; finally it accords with the Josephson plasma energy for the sufficiently high barrier.
Thus, the crossover from the dipole mode to the Josephson plasma mode is confirmed in the case of a realistic double-well potential.

In order to elucidate the role of the anomalous tunneling in the crossover, we next estimate half-width $\Delta\varepsilon$ of the peak of the anomalous tunneling for a Gaussian potential barrier.
In case of a rectangular potential barrier, Kagan {\it et al}. have approximately obtained an analytical expression for the transmission coefficient.~\cite{rf:antun}
Using the expression, one can calculate the half-width,
  \begin{eqnarray}
  \frac{\Delta\varepsilon}{\mu}\simeq 
  \frac{2\sqrt{2}\,{\rm e}^{-\kappa_0 d}}{\kappa_0 \xi},\label{eq:half_w}
  \end{eqnarray}
where
  \begin{eqnarray}
  \kappa_0 = \frac{1}{d}\int_{-\frac{d}{2}}^{\frac{d}{2}}dx
  \sqrt{\frac{2m}{\hbar^2}(V_{\rm barrier}(x)-\mu)}.\label{eq:t_ac}
  \end{eqnarray}
In the region $|x|<\frac{d}{2}$, the potential barrier $V_{\rm barrier}(x)$ is larger than the chemical potential; this means that the points $|x|=\frac{d}{2}$ are the classical turning points.
We estimate the half-width $\Delta\varepsilon$ for a Gaussian potential barrier by means of the expression of eqs. \!(\ref{eq:half_w}) and (\ref{eq:t_ac}).

Triangles in Fig. \ref{fig:harmo} represent the half-width $\Delta\varepsilon$ of the peak of the transmission coefficient.
When the dipole mode energy is much smaller than the half-width, namely $\hbar\omega_x \ll \Delta\varepsilon$, the anomalous tunneling is effective for the lowest excitation energy.
Accordingly, the lowest excitation is hardly affected by the potential barrier.
When $\hbar\omega_x \sim \Delta\varepsilon$, the lowest excitation energy remarkably changes because the excitation begins to perceive the potential barrier.
When $\hbar\omega_x \gg \Delta\varepsilon$, the anomalous tunneling is no longer effective; consequently, the lowest excitation becomes the Josephson plasma mode.
Therefore, the interpretation of the crossover discussed in the previous section is also appropriate for the double-well potential of eq. \!(\ref{eq:realpote}).
\section{Conclusions}\label{sec:conc}
In summary, we have studied collective excitations of a condensate in a double-well potential.
We have analytically solved the Bogoliubov equations with a box-shaped double-well potential, and analyzed the excitation spectrum. 
In the lowest excitation, it has been found that the crossover from the dipole mode to the Josephson plasma mode occurs as the potential barrier separating the condensate becomes strong.
The crossover is dominated by the anomalous tunneling behavior of the Bogoliubov excitations.
We have numerically calculated the lowest excitation energy and the Josephson plasma energy for a condensate in a realistic double-well potential.
We have confirmed by numerical calculation that the mechanism of the crossover is valid also in the case of more realistic potential.

While only condensates at $T=0$ has been considered in our calculations, the Josephson plasma oscillation of condensates in finite temperature is known to exhibit dissipative behaviors due to the presence of the thermal depletion~\cite{rf:zwerg}.
It will be interesting to study the excitations of the condensates at finite temperature in the double-well potential, and to discuss the relation between the damping of the collective oscillation and the anomalous tunneling of the Bogoliubov excitations.

\section*{Acknowledgment}
We would like to thank S. Tsuchiya, T. Kimura and T. Nikuni for useful discussions.
This work is partly supported by a Grant for The 21st Century COE Program (Physics of Self-organization Systems) at Waseda University from the Ministry of Education, Sports, Culture, Science, and Technology of Japan.
I.D. is supported by Grant-in-Aid for JSPS fellows.
%


\begin{thebibliography}{99} 
\bibitem{rf:interfere}
M. R. Andrews, C. G. Townsend, H.-J. Miesner, D. S. Durfee, D. M. Kurn and W. Ketterle: Science {\bf 275} (1997) 637.

\bibitem{rf:nondestructive}
M. Saba, T. A. Pasquini, C. Sanner, Y. Shin, W. Ketterle and D. E. Pritchard: Science {\bf 307} (2005) 1945.

\bibitem{rf:joseph_origin}
B. D. Josephson: Phys. Lett. {\bf 1} (1962) 251.

\bibitem{rf:smerzi}
A. Smerzi, S. Fantoni, S. Giovanazzi, and S. R. Shenoy: Phys. Rev. Lett. {\bf 79} (1997) 4950.

\bibitem{rf:zwerg}
F. Meier and W. Zwerger: Phys. Rev. A {\bf 64} (2001) 033610.

\bibitem{rf:BEC}
L. P. Pitaevlkii and S. Stringari: {\it Bose-Einstein Condensation} (Oxford University Press, Oxford, 2003).

\bibitem{rf:Leg}
A. J. Leggett: Rev. Mod. Phys. {\bf 73} (2001) 307.

\bibitem{rf:JPST}
M. Albiez, R. Gati, J. F\o lling, S. Hunsmann, M. Cristiani and M. K. Oberthaler: Phys. Rev. Lett. {\bf 95} (2005) 010402.

\bibitem{rf:parao}
G.-S. Paraoanu, S. Kohler, F. Sols and A. J. Leggett: J. Phys. B: At. Mol. Opt. Phys. {\bf 34} (2001) 4689.

\bibitem{rf:numel_dw}
L. Salasnich, A. Parola and L. Reatto: Phys. Rev. A {\bf 60} (1999) 4171.

\bibitem{rf:antun}
Yu. Kagan, D. L. Kovrizhin and L. A. Maksimov: Phys. Rev. Lett. {\bf 90} (2003) 130402.

\bibitem{rf:box}
T. P. Meyrath, F. Schreck, J. L. Hanssen, C.-S. Chuu and M. G. Raizen: Phys Rev. A {\bf 71} (2005) 041604(R).

\bibitem{rf:1dime}
M. Olshanii: Phys. Rev. Lett. {\bf 81} (1998) 938.

\bibitem{rf:KP}
I. Danshita, S. Kurihara and S. Tsuchiya: cond-mat/0506266.

\bibitem{rf:kovri}
D. L. Kovrizhin: Phys. Lett. A {\bf 287} (2001) 392.

\end{thebibliography}
\end{document}